\documentclass[a4paper,american,floatfix,pdftex,superscriptaddress,twoside,%
aip,rsi,%
citeautoscript,%
reprint,twocolumn,
]{revtex4-2}%

\usepackage{amsfonts,amsmath,amssymb}
\usepackage[T1]{fontenc}
\usepackage{graphicx}%
\usepackage[utf8]{inputenc}
\usepackage{microtype}
\usepackage{xspace}
\usepackage{hyperref, hypernat}
\usepackage[todos]{CMImacros}

\graphicspath{{./}} 
\hypersetup{citebordercolor=yellow,linkbordercolor=red,urlbordercolor=blue} %

\newcommand*{\pyrrolei}{\ensuremath{\pyrrole^+}\xspace}

\makeatletter%
\renewcommand*{\@fnsymbol}[1]{\ensuremath{\ifcase#1\or \|\or \mathsection\or \ddagger\or *\or **\or
      \mathparagraph\or \dagger\or \dagger\dagger \or \ddagger\ddagger \else\@ctrerr\fi}}%
\makeatother

\newcommand{\cfeldesy}{\affiliation{Center for Free-Electron Laser Science CFEL, \mbox{Deutsches
         Elektronen-Synchrotron DESY}, Notkestr.~85, 22607~Hamburg, Germany}}%
\newcommand{\jlu}{\affiliation{Institute of Atomic and Molecular Physics, Jilin University,
      Changchun 130012, China}}%
\newcommand{\uhhcui}{\affiliation{Center for Ultrafast Imaging, Universität Hamburg, Luruper
      Chaussee 149, 22761 Hamburg, Germany}}%
\newcommand{\uhhphys}{\affiliation{Department of Physics, Universität Hamburg, Luruper Chaussee 149,
      22761 Hamburg, Germany}}%
\newcommand{\desy}{\affiliation{Deutsches Elektronen-Synchrotron DESY, Notkestr. 85, 22607 Hamburg,
      Germany}}%
\newcommand{\xfel}{\affiliation{European XFEL, Holzkoppel 4, 22869 Schenefeld, Germany}}%
\newcommand{\lund}{\affiliation{Department of Physics, Lund University, 22100 Lund, Sweden}}%
\newcommand{\kansas}{\affiliation{J. R. Macdonald Laboratory, Department of Physics, Kansas State
      University, Manhattan, Kansas 66506, USA}}%
\newcommand{\cnrs}{\affiliation{Sorbonne Université, CNRS, Laboratoire de Chimie Physique-Matière et
      Rayonnement, LCPMR, F-75005 Paris, France}}%
\newcommand{\freie}{\affiliation{Physical Chemistry, Freie Universität Berlin, Arnimallee 22, 14195
      Berlin, Germany}}%
\newcommand{\cnr}{\affiliation{ISM-CNR, Istituto Struttura della Materia, LD2 Unit, Basovizza Area
      Science Park, 34149 Trieste, Italy}}%
\newcommand{\scpa}{\affiliation{Elettra-Sincrotrone Trieste S.C.P.A., 34149 Basovizza, Trieste,
      Italy}}%
\newcommand{\guf}{\affiliation{Institut für Kernphysik, Goethe-Universität Frankfurt,
      Max-von-Laue-Str. 1, 60438 Frankfurt am Main, Germany}}
\newcommand{\fhi}{\affiliation{Molecular Physics, Fritz-Haber-Institut der Max-Planck-Gesellschaft,
    Faradayweg 4-6, 14195 Berlin, Germany}}

\newcommand{\run}{\altaffiliation[Current address: ]{Radboud
      University, Institute for Molecules and Materials, Heijendaalseweg 135, 6525 AJ Nijmegen, The
      Netherlands}}

\newcommand{\vins}{\altaffiliation[Current address: ]{Vinča Institute of Nuclear Sciences, National
    Institute of the Republic of Serbia, University of Belgrade, 12-14 Mike Petrovića Alasa, 11351
    Vinča, Begrade, Serbia}}

\newcommand{\stemail}{\email[]{sebastian.trippel@cfel.de}}
\newcommand{\jkemail}{\email[]{jochen.kuepper@cfel.de}}%
\newcommand{\cmiweb}{\homepage{https://www.controlled-molecule-imaging.org}}%

\begin{document}
\onecolumngrid%
\title{Controlled molecule injector for cold, dense, and pure molecular beams \\
   at the European x-ray free-electron laser}%
\author{Lanhai~He}\cfeldesy\jlu%
\author{Melby~Johny}\cfeldesy\uhhcui\uhhphys%
\author{Thomas~Kierspel}\cfeldesy\uhhcui\uhhphys%
\author{Karol~Długołęcki}\cfeldesy%
\author{Sadia~Bari}\desy%
\author{Rebecca~Boll}\xfel%
\author{Hubertus~Bromberger}\cfeldesy%
\author{Marcello~Coreno}\cnr\scpa%
\author{Alberto~De~Fanis}\xfel%
\author{Michele~Di~Fraia}\scpa%
\author{Benjamin~Erk}\desy%
\author{Mathieu~Gisselbrecht}\lund%
\author{Patrik~Grychtol}\xfel%
\author{Per~Eng-Johnsson}\lund%
\author{Tommaso~Mazza}\xfel%
\author{Jolijn~Onvlee}\run\cfeldesy\uhhcui%
\author{Yevheniy~Ovcharenko}\xfel%
\author{Jovana~Petrovic}\vins\cfeldesy%
\author{Nils~Rennhack}\xfel%
\author{Daniel~E.~Rivas}\xfel%
\author{Artem~Rudenko}\kansas%
\author{Eckart~Rühl}\freie%
\author{Lucas~Schwob}\desy%
\author{Marc~Simon}\cnrs%
\author{Florian~Trinter}\fhi\guf
\author{Sergey~Usenko}\xfel%
\author{Joss~Wiese}\cfeldesy%
\author{Michael~Meyer}\xfel%
\author{Sebastian~Trippel}\stemail\cfeldesy\uhhcui
\author{Jochen~Küpper}\jkemail\cmiweb\cfeldesy\uhhcui\uhhphys
\begin{abstract}\noindent
   A permanently available molecular-beam injection setup for controlled molecules (COMO) was
   installed and commissioned at the small quantum systems (SQS) instrument at the European x-ray
   free-electron laser (EuXFEL). A $b$-type electrostatic deflector allows for pure state-, size-,
   and isomer-selected samples of polar molecules and clusters. The source provides a rotationally
   cold ($T\approx1$~K) and dense ($\rho\approx10^8$~cm$^{-3}$) molecular beam with pulse durations
   up to 100~\us generated by a new version of the Even-Lavie valve. Here, a performance overview of
   the COMO setup is presented along with characterization experiments performed both, with an
   optical laser at the Center for Free-Electron-Laser Science and with x-rays at EuXFEL under
   burst-mode operation. COMO was designed to be attached to different instruments at the EuXFEL, in
   particular at the small quantum systems (SQS) and single particles, clusters, and biomolecules
   (SPB) instruments. This advanced controlled-molecules injection setup enables XFEL studies using
   highly defined samples with soft and hard x-ray FEL radiation for applications ranging from
   atomic, molecular, and cluster physics to elementary processes in chemistry and biology.
\end{abstract}
\maketitle

X-ray free-electron lasers (XFELs) deliver x-ray flashes with unprecedentedly high intensities in
combination with ultra-short pulse durations down to the attosecond
regime~\cite{McNeil:NatPhoton4:814, Emma:NatPhoton4:641, Altarelli:NIMB269:2845,
   Ishikawa:NatPhoton6:540, Milne:AP7:720, Kang:NatPhoton11:708, Allaria:NatPhoton7:913,
   Hartmann:NatPhoton12:215, Decking:NatPhoton14:391}. With the development and enormous scientific
success of these large-scale light sources such as the free-electron laser in Hamburg (FLASH), the
linac coherent light source (LCLS), the free-electron laser radiation for multidisciplinary
investigations (FERMI), the SPring-8 Angstrom compact free-electron laser (SACLA), the x-ray
free-electron laser at the Paul Scherrer Institute (SwissFEL), the European x-ray free-electron
laser (EuXFEL), and the Shanghai soft x-ray free-electron laser (SXFEL), a huge demand for user
experiments arose~\cite{Strueder:NIMA614:483, Zhao:ApplSci7:607, Young:JPB51:032003, Erk:JSR25:1529,
   Kuster:JSR28:576}. In this context, applications with very cold and very pure gas-phase molecular
beams are highly desirable, especially for experiments performed in the field of atomic, molecular,
and optical (AMO) sciences~\cite{Young:Nature466:56, Rudek:NatPhoton6:858, Feldhaus:JPB46:164002,
   Eichmann:Science369:1630, Li:PRL127:093202, Fehre:PRL127:103201, Lee:NatComm12:6107}.

Molecular beam methods were established to obtain fundamental insights into the mechanisms and
dynamics of elementary molecular and chemical processes. Starting with the first
experiments~\cite{Dunoyer:ComptRend152:592, Stern:ZP39:751}, molecular beam methods were further
developed and refined over the last hundred years and led to tremendous advances in the scientific
understanding and manipulation of small molecules in the gas phase. Highlights include the discovery
of the intrinsic spin~\cite{Stern:PZ23:476}, the discovery of nuclear magnetic
moments~\cite{Frisch:ZP85:41933gp, Rabi:PR55:526}, the investigation of chemical reaction
dynamics~\cite{Herschbach:DFS33:149}, and the invention of the MASER~\cite{Gordon:PR99:1264}. In
particular, in recent years the control of molecules with external fields opened the door to
completely new experiments such as scattering experiments at very small relative energies where
quantum effects predominate~\cite{Meerakker:NatPhys4:595, Meerakker:CR112:4828, Chang:IRPC34:557,
   Zastrow:NatChem6:216, Henson:Science338:234, Jankunas:JPCA118:3875}. Using the electrostatic
deflector, pure beams of state-, size-, and isomer-selected samples of polar molecules and clusters
were realized~\cite{Filsinger:PRL100:133003, Filsinger:ACIE48:6900, Trippel:RSI89:096110,
   Chang:Science342:98, Teschmit:ACIE57:13775, Onvlee:NatComm13:7462}. Furthermore, due to their
rotational temperature on the order of 0.1~K, these pure samples enable very strong laser alignment
and mixed field orientation~\cite{Stapelfeldt:RMP75:543, Holmegaard:PRL102:023001,
   Ghafur:NatPhys5:289, Mullins:NatComm13:1431, Trippel:JCP148:101103}, which can be exploited in
the recording of so-called molecular movies, \eg, time-resolved diffractive imaging in the molecular
frame to directly determine the molecular structure and dynamics by mathematical transformations
from the measured diffraction patterns~\cite{Spence:PRL92:198102, Hensley:PRL109:133202,
   Kuepper:PRL112:083002, Barty:ARPC64:415, Zhang:NatComm12:5441, Centurion:ARPC73:21}.

Facilities like FLASH, EuXFEL, or LCLS-II, with repetition rates up to 4.5~MHz, provide ultrashort
x-ray pulses with brilliances that are a billion times higher than that of the best conventional
x-ray radiation sources~\cite{Beye:EPJP138:193, SLAC:New-Science-LCLS-II:2016,
  Altarelli:NIMB269:2845, Decking:NatPhoton14:391}. To match the burst-mode operation of the light
pulses at EuXFEL, a cold, pure, and pulsed molecular beam with a duration on the order of 400~\us
and a fundamental repetition rate of 10~Hz is desirable. Therefore, the COMO setup was developed as
a permanently available extension of endstations at the SQS and SPB instruments at EuXFEL. COMO is
designed based on the concept of a supersonic-molecular-beam source combined with the electrostatic
$b$-type deflector~\cite{Kuepper:PRL112:083002, Kienitz:JCP147:024304}.

Here, we present an overview of the design and the capabilities of the permanently available
endstation-extension COMO as demonstrated at the SQS instrument of the EuXFEL, where COMO serves as
a source for species-selected, cold, dense, and ultra-pure molecular beams.

\section{Apparatus description}
\subsection{Vacuum system}
The vacuum setup, shown schematically in the lower right corner of \autoref{fig:setup}, consists of
two stainless steel chambers, highlighted by the blue and purple areas, housing the molecular beam
source and the electrostatic $b$-type deflector, respectively. The source chamber is pumped with two
turbomolecular pumps (Pfeiffer Vacuum HiPace 2300), resulting in a pumping speed on the order of
4000~L/s for helium. The deflector chamber is pumped with two turbomolecular pumps (Pfeiffer Vacuum
HiPace 700), resulting in a pumping speed on the order of 1400~L/s. The typical pressure in the
source- and deflector chamber when operating the molecular beam valve at 10~Hz is $\ordsim10^{-7}$
and $\ordsim10^{-8}$~mbar, respectively. Both chambers can be separated with a manual DN250 CF gate
valve (VAT 10848-CE01-0008) to allow for sample changes in the source chamber without breaking the
vacuum of the deflector chamber and the endstation. The source chamber is equiped with a quick
access door (Pfeiffer Vacuum 420KTU250) on a DN250 CF flange to allow for fast sample replacement in
the valve's sample container~\cite{Even:EPJTI:2:17}. The source and deflector chambers are mounted
on rails attached to a stainless steel platform to facilitate easy connection of the entire setup to
the endstation. The total mass of the molecular beam setup including the pumps is $\ordsim500$~kg.
The interface between COMO and an endstation requires a DN250 CF port with a cylindrical shaped
space with a length of 13.5~cm and a diameter of 20.4~cm inside the endstation for the inverse
skimmer pot.

\subsection{Molecular beam setup}
\begin{figure*}
   \includegraphics[width=\linewidth]{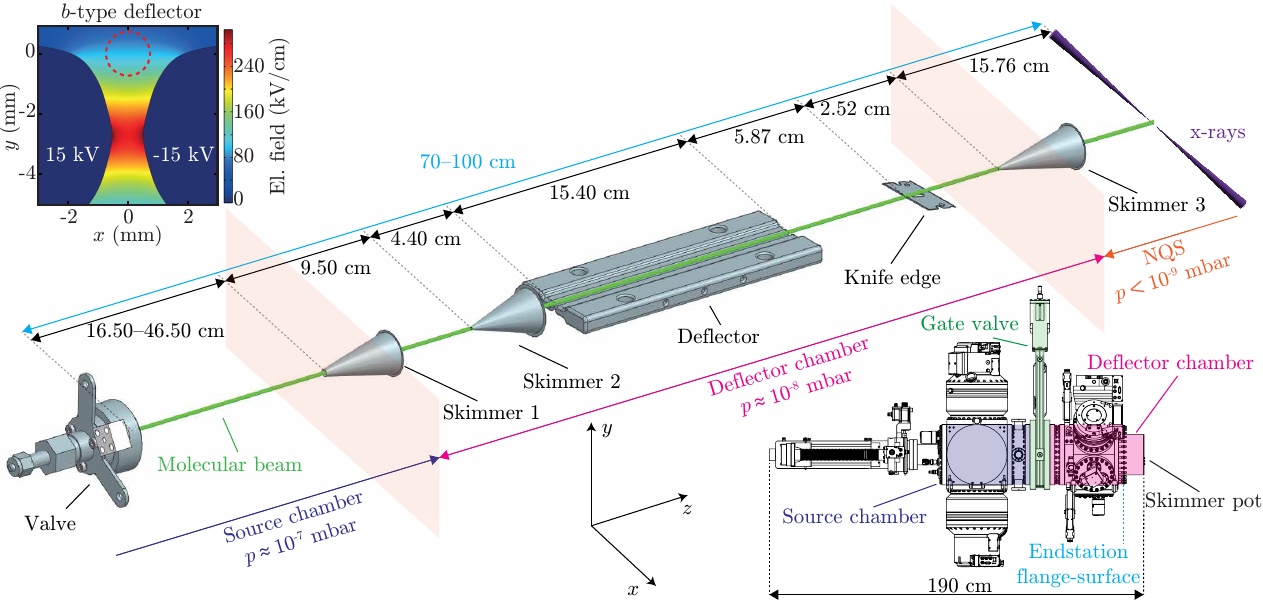}%
   \caption{Schematic of the COMO setup with its main constituents of a pulsed valve and the
     electrostatic $b$-type deflector. X-rays and possibly further optical laser beams cross the
     molecular beam in the interaction region of, for example, the NQS endstation at the SQS
     instrument. Semi-transparent planes indicate differential pumping sections at the first and
     third skimmer, respectively. The indicated pressures are typical values during operation with
     long pulses at 10~Hz. The complete COMO vacuum setup is sketched in the lower right corner. The
     flange surface of an endstation is indicated by the light blue dashed line. The skimmer pot
     ranges into the endstation. The upper left inset depicts the shape and electric fields of the
     $b$-type deflector geometry. The numerically calculated electric field strengths are depicted
     in color coding. The dashed circle indicates the typical size and position of the undeflected
     molecular beam.}
   \label{fig:setup}
\end{figure*}

A schematic of the COMO molecular beam setup is shown in \autoref{fig:setup}.
A commercially available long-pulse version of an Even-Lavie (EL) valve~\cite{Even:EPJTI:2:17} is
used to deliver the molecular beam by a supersonic expansion of helium with a trace of molecules
into the vacuum. This version of the valve produces molecular beam pulse durations up to 100~\us by
utilizing two current pulses in close succession provided by a modified driver unit. The valve is
situated on a 3D manipulator (VAB PM12-300-S2EC) with travel ranges of 1.25~cm, 1.25~cm, and 30~cm
in $x$, $y$, and $z$ directions, respectively. The valve body can be heated up to \celsius{250} to
allow for vapor pressure control of the sample under investigation. Helium is usually used as the
carrier gas with a stagnation pressure on the order of tens of bars, up to 100~bar, to generate a
cold molecular beam. At EuXFEL, the valve is operated at 10~Hz, matching the x-ray burst rate. A
fast beam flux monitor (MBE-Komponenten BFM 40-150) mounted on a DN40~CF linear feedthrough with a
linear travel of 15~cm is used to monitor the temporal profile of the molecular beam. The beam flux
monitor is controlled by a modified pressure gauge controller (JEVATEC VCU-B0) that allows for
measuring the collector current. The collector current is transimpedance amplified with a current
amplifier (FEMTO DHPCA-100), which in turn is connected to a digitizer to display and record the
temporal profile of the molecular beam.

All of the components used to shape the molecular beam are mounted on motorized translation stages
and the individual positions can be remotely controlled in planes perpendicular to the molecular
beam propagation direction. The first skimmer (Beam Dynamics 50.8, $\varnothing=3$~mm) is used for
differential pumping. The distance between the first skimmer and the valve can be adjusted between
16.5 and 46.5~cm by moving the valve along the molecular beam propagation direction. The first
skimmer is mounted on a home-built flange with a built-in 2D translation stage to adjust the skimmer
position in the $x$ and $y$ directions using two stepper motors (Thorlabs, DRV014) outside vacuum
and linear feedthroughs. The second skimmer (Beam Dynamics 40.5, $\varnothing=1.5$~mm) is placed
9.5~cm behind the first skimmer tip for further collimation of the beam. An electrostatic $b$-type
deflector~\cite{Kienitz:JCP147:024304}, which disperses the molecules in the beam with respect to
their quantum states~\cite{Chang:IRPC34:557, Filsinger:JCP131:064309} is located 4.4~cm downstream
the tip of the second skimmer. Both electrodes of the electrostatic deflector are connected through
two high-voltage feedthroughs (Pfeiffer, 420XST040-30-30-1) allowing for voltages differences up to
$60$~kV between the electrodes. The unit of the second skimmer and the electrostatic deflector are
mounted on a 2D translation stage to adjust their positions simultaneously in the $x$ and $y$
directions using two stepper motors (Thorlabs DRV014) and linear feedthroughs. This translation
stage is mounted on the back side of the same flange as the translation stage for skimmer 1. The
upper left inset of \autoref{fig:setup} depicts the shape and electric fields of the $b$-type
deflector geometry. The numerically calculated electric field strengths are depicted in color
coding. The dashed circle indicates the typical size and position of the undeflected molecular beam.
The dispersed molecular beam is then cut by a vertically adjustable knife edge placed 5.87~cm
downstream of the exit of the deflector. This allows for both, improved sample separation and higher
column density of the molecular beam~\cite{Trippel:RSI89:096110}. The orientation of the knife edge
can be controlled by a motorized rotation stage (Smaract SR-1908) to ensure a molecular beam-cut
parallel to the x-ray propagation direction. The vertical position of the knife edge can be adjusted
by a linear in-vacuum piezo stage (Smaract SLC 1750) with a travel range of 3.1~cm. The typical
pressure in the deflector chamber is on the order of $10^{-8}$~mbar in operation. The molecular beam
is further skimmed by a third conical skimmer (Beam Dynamics 50.8, $\varnothing=1.5$~mm) placed
2.52~cm downstream of the knife edge, providing differential pumping against the interaction
chamber. This third skimmer is again mounted on a 2D translation stage, with the position controlled
by two linear in-vacuum piezo stages (Smaract SLC 1750). Behind the third skimmer, the molecular
beam enters an endstation compatible with COMO, \eg, the nano-size quantum systems (NQS), or the
reaction microscope (REMI) experimental stations at the SQS instrument at
EuXFEL~\cite{Tschentscher:ApplSci, Decking:NatPhoton14:391}. The distance between the tip of the
third skimmer and the interaction region with the x-rays is 15.76~cm in case of the NQS setup.

\section{Characterization of the setup}
\subsection{Molecular beam temporal profile}
To characterize the performance of the COMO setup in-house before its delivery to EuXFEL, it was
combined with a time-of-flight mass spectrometer (TOF-MS, Jordan C-677) and tested with a
Ti:Sapphire femtosecond laser system (Coherent Astrella) to strong-field-ionize the sample. The
laser pulses with a duration of $\mathrm{FWHM}_I\sim30$~fs and a wavelength centered at 800~nm were
focused to $\mathrm{FWHM}_I\approx50$~\um and directed perpendicular to the molecular beam along the
negative $x$ direction to ionize molecules in the extraction region of the TOF-MS. The TOF-MS had a
distance of 17.6~cm with respect to the last skimmer tip along the $z$ direction, reflecting the
geometry at the EuXFEL SQS instrument. The peak intensity of the laser pulse was
$I_0 \sim 6\times10^{13}~\Wpcmcm$. We used pyrrole as a sample to characterize the temporal and
spatial profiles of the molecular beam~\cite{Johny:CPL721:149}. Pyrrole (Sigma-Aldrich Chemie GmbH,
131709) was placed in the sample reservoir of the valve and heated up to \celsius{70}, which
resulted in a pyrrole vapor pressure on the order of 100~mbar~\cite{Eon:JCED16:408}. Helium was used
as the carrier gas at a stagnation pressure of 20~bar.

\autoref[(a)]{fig:molbeam_profiles} shows the temporal density profiles of molecular-beam pulses
obtained, gating the signal on the parent cation of pyrrole (\mq{67}), for both, a long-pulse
(violet) and a standard EL valve (green).
\begin{figure}
   \includegraphics[width=\linewidth]{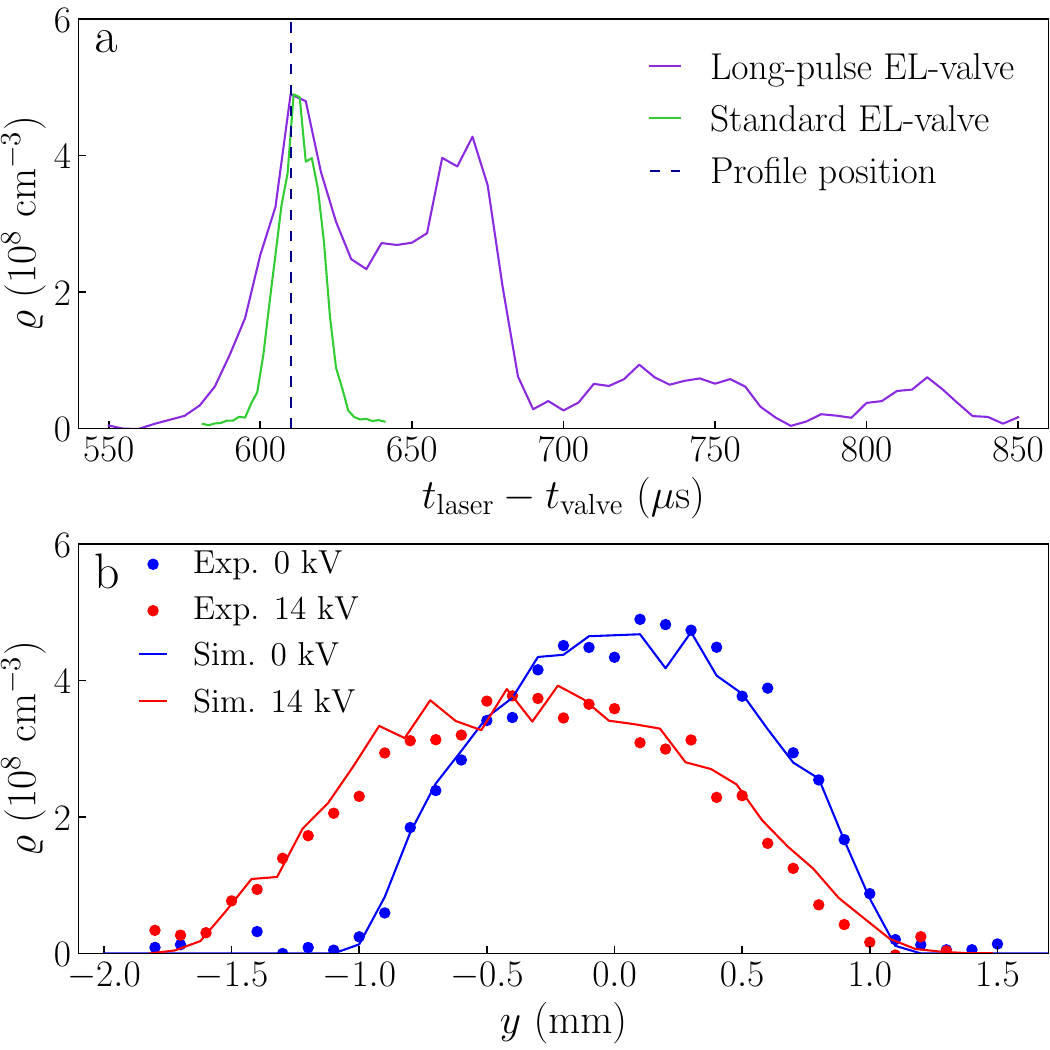}%
   \caption{(a) Temporal molecular-beam-density profiles provided by a long-pulse EL valve (violet)
     and a normal EL valve (green) obtained from \pyrrolei. The density of the normal EL valve is
     not to scale. The dashed line indicates the relative timing where the spatial molecular beam
     density profiles were measured. (b) Spatial molecular-beam-density profiles for the deflector
     switched off (blue) and on (red) at a stagnation pressure of 20~bar and a probe time of
     610~\us.}
   \label{fig:molbeam_profiles}
\end{figure}
A pulse duration of $\tau\sim100$~\us is observed for the long-pulse valve in contrast to
$\tau\sim25$~\us in the case of the normal EL valve. In addition, for the long-pulse EL valve, two
maxima can be identified during the pulse, resulting from the two current pulses of the driver unit.
Furthermore, post pulses were detected at delays between 700--850~\us, which are attributed to the
bouncing of the plunger in the EL valve driven by the two current pulses. The post pulses are
expected to gradually disappear with increasing stagnation
pressure~\cite{Christen:JCP139:154202}. The absolute density for the long-pulse EL valve was
determined by a laser-pulse-energy scan as described below. The density of the normal EL valve is
not to scale, but from a signal strengths comparison, we assume similar densities as obtained for
the long-pulse EL valve.

\subsection{Molecular beam spatial profiles}
To characterize the separation of pyrrole from the direct molecular beam, measurements were carried
out with and without voltages on the deflector. \autoref[(b)]{fig:molbeam_profiles} shows the
measured (dots) and simulated~\cite{Chang:IRPC34:557} (lines) profiles for the direct (deflector
off, blue) and the $14$~kV ($\pm 7$~kV) deflected (red) molecular beams. Both profiles were obtained
by scanning the molecular beam along the $y$ direction making use of the motorized translation
stages to move the beam with respect to the laser propagation axis. The parent ion yields were
obtained by integrating over the \pyrrolei peak in the TOF-MS. The timing of the laser pulses with
respect to the molecular-beam-valve trigger was 610~\us, as indicated by the vertical dashed line in
\autoref[(a)]{fig:molbeam_profiles}. Comparing both profiles, a clear shift toward negative $y$
values is observed when the deflector is switched on. This is expected, as all quantum states of
pyrrole are strong-field seeking at the relevant electric field strengths in the
deflector~\cite{Johny:CPL721:149}. The solid lines in \autoref[(b)]{fig:molbeam_profiles} are
simulated pyrrole beam profiles obtained from Monte-Carlo trajectory calculations that take into
account the geometrical constraints of the mechanical apertures and the rotational temperature
$T_\mathrm{rot}$ of the molecular beam~\cite{Chang:CPC185:339, Johny:CPL721:149}. The experimental
data of pyrrole's deflection profiles match well with the simulated results, which assume an initial
temperature of $T_\mathrm{rot}=1$~K for the molecular beam entering the deflector. Furthermore, the
results are comparable to those obtained in earlier measurements~\cite{Johny:CPL721:149,
  Johny:PCCP26:13118}.

\subsection{Molecular beam density determination}
The pyrrole molecular beam sample densities were estimated based on a strong-field ionization
model~\cite{Hankin:PRA64:013405, Wiese:NJP21:083011, Johny:PCCP26:13118}: The asymptotic slope of an
integral ionization signal with respect to the natural logarithm of peak intensity can be expressed
as $S = \varrho\pi R^{2}D\alpha$. Here, $\varrho$ represents the sample density, $R$ the $I_0/e$
radius of the transverse laser intensity distribution, $D$ the length of the focal volume in the
molecular beam, and $\alpha$ the detection efficiency.
\begin{figure}
   \includegraphics[width=\linewidth]{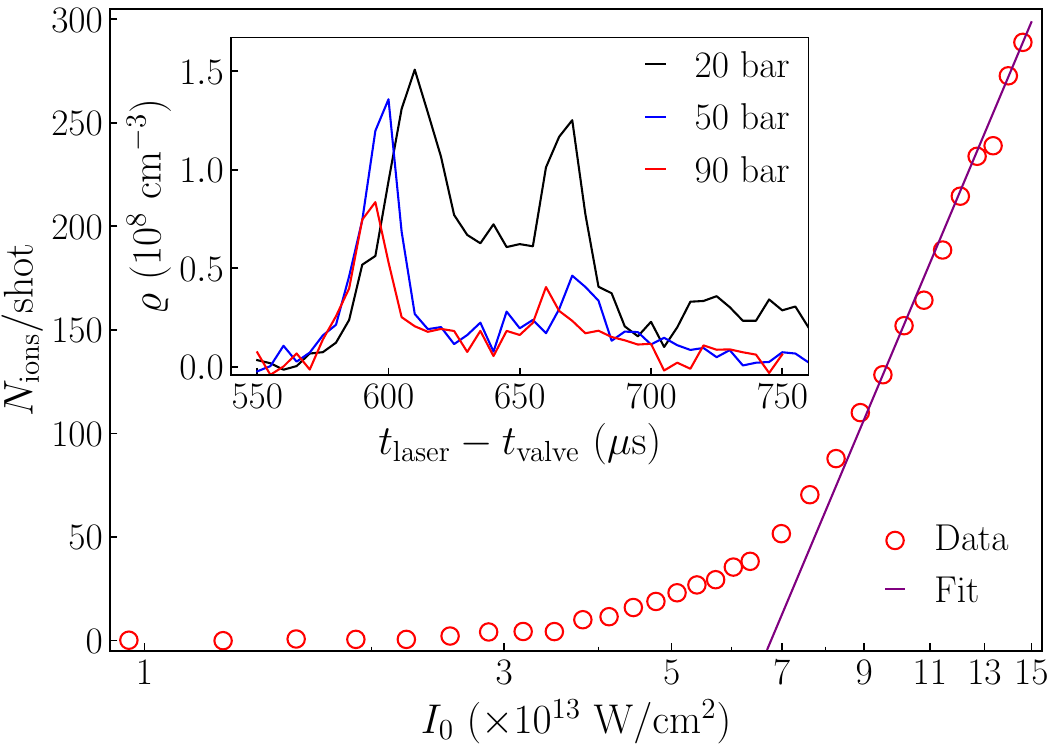}
   \caption{Number of pyrrole ions per laser pulse (red circles) as a function of the laser
     peak-intensity logarithm $\ln(I_0)$ for a stagnation pressure of 90~bar. The purple line
     indicates the asymptotic-slope fit, see text for details. The inset shows temporal
     molecular-beam-density profiles under different valve stagnation pressures of 20~bar (black),
     50~bar (blue), and 90~bar (red).}
   \label{fig:ion-yield}
\end{figure}
The detected number of pyrrole ions per laser pulse as a function of the logarithm of the laser peak
intensity $\ln(I_0)$ is shown in \autoref{fig:ion-yield}. The vertical laser position, molecular
beam timing, and stagnation pressure was $y=0$~mm, $t=600$~\us, and 90~bar, respectively. A typical
increase and saturation behavior of the ion yield with increasing laser intensity is observed. The
purple curve depicts a fit to the asymptotic slope. From the fit, a saturation onset of
$I_0=(6.8\pm 1.1_\text{syst})\times10^{13}~\Wpcmcm$ was deduced. A sample density of
$\varrho=(8.4 \pm 4.2_\text{syst})\times10^7~\mathrm{cm}^{-3}$ was obtained, taking into account the
measured laser-beam waist of $\omega_0=2\sigma = 52(2)$~\um, the measured molecular beam diameter of
$D=2.0(3)$~mm, and an estimated detection efficiency of 50\%~\cite{Fehre:RSI89:045112}. The measured
molecular beam density is consistent with typical previously reported
values~\cite{Kuepper:PRL112:083002, Wiese:NJP21:083011, Johny:PCCP26:13118} obtained with
an EL valve. The inset in \autoref{fig:ion-yield} shows the temporal profiles of the pyrrole
molecular beam for stagnation pressures of 20, 50, and 90~bar, respectively. A decrease in pulse
duration is observed with increasing stagnation pressure accompanied by a reduction in peak density.
Currently, this limits the long-pulse operation to samples where sufficient cooling is achieved at
20~bar stagnation pressure. Improvements in the valve-driving-electronics parameters are foreseen
for samples of controlled and species-selected molecular clusters~\cite{Johny:CPL721:149,
   Trippel:PRA86:033202, Onvlee:NatComm13:7462} or applications of strong molecular
orientation\cite{Holmegaard:PRL102:023001, Holmegaard:NatPhys6:428}, which require higher stagnation
pressures for efficient formation or cooling.

\subsection{Characterization experiments with x-rays}
\begin{figure}
   \includegraphics[width=\linewidth]{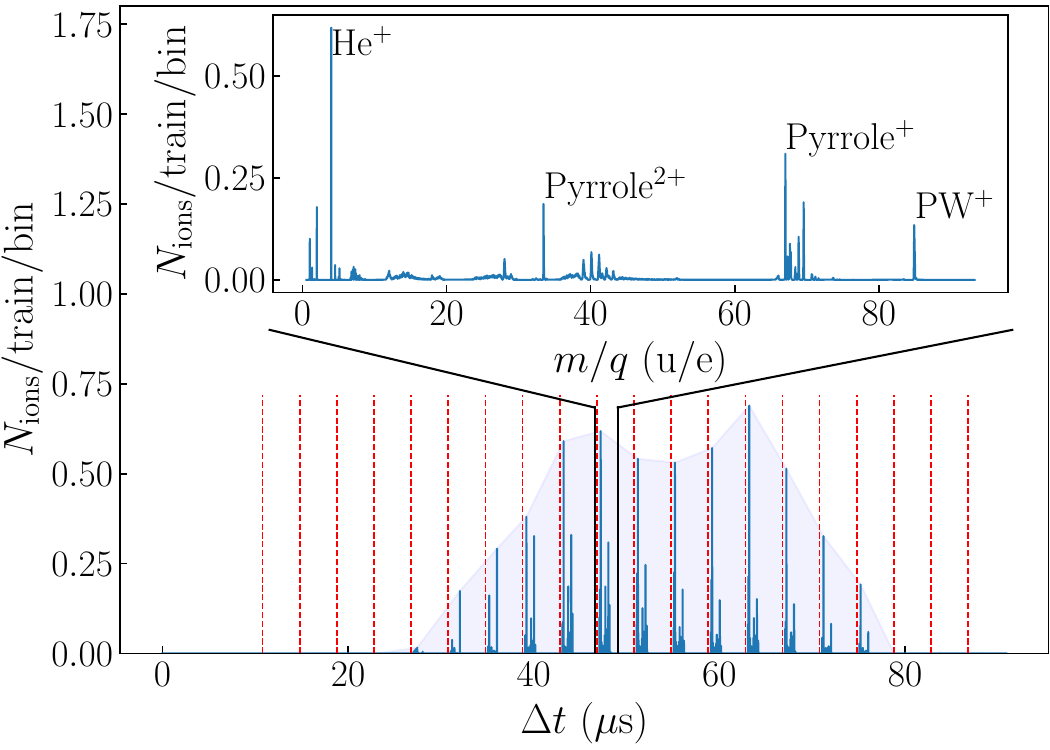}
   \caption{Ion-yield spectrum induced by the combination of FEL- and 1030~nm optical laser pulse
     trains at a base frequency of 10~Hz with 20~FEL/1030~nm~pulses per train at an intra-train
     repetition rate of 250~kHz and a photon energy of 600~eV. $\Delta t$ is the relative timing
     between the molecular beam arrival time and the train of x-ray pulses. Red vertical dotted
     lines indicate the arrival of each FEL/1030~nm pulse in the train. The blue shaded area
     indicates the temporal envelope of the molecular beam density. One bin corresponds to 1.136~ns.
     The inset is a zoomed-in TOF-MS spectrum corresponding to one of the FEL/1030~nm pulses with
     the main peaks labeled.}
   \label{fig:tof}
\end{figure}

The COMO setup was assembled and further tested with the nano-sized quantum systems (NQS)
experimental station at the SQS instrument at EuXFEL. A pure pyrrole-water (PW) heterodimer
sample~\cite{Johny:CPL721:149, Johny:PCCP26:13118} was investigated. The molecular beam was produced
using a stagnation pressure of 80~bar of helium with traces of water and pyrrole added. The driver
settings were similar to the ones used during the characterization measurements with the optical
laser described above. These conditions resulted in a similar spatial molecular beam profile and a
purity of pyrrole-water heterodimer as published previously~\cite{Johny:CPL721:149,
  Johny:PCCP26:13118} and also shown in \autoref[(b)]{fig:molbeam_profiles}. The EuXFEL and the
optical laser were operated in burst mode at a base frequency of 10~Hz with 20~FEL pulses,
synchronized to the same number of 1030~nm optical laser pulses, per train at an intra-train
repetition rate of 250~kHz and a photon energy of 600~eV. \autoref{fig:tof} depicts the typical
ion-yield spectrum as a function of the relative timing $\Delta t$ between the molecular beam
arrival time and the train of combined x-ray- and 1030~nm optical laser pulses.
Vertical red dotted lines indicate the timing of each specific x-ray pulse in the train. The signal
envelope as depicted by the blue shaded area indicates the overall temporal profile of the molecular
beam in the deflected part. The pulse duration of the molecular beam was on the order of 40~\us
during the commissioning time. The temporal molecular-beam profile is fully covered by the
20~FEL/1030~nm~pulses and the specific molecular beam TOF spectra and the background are detected
simultaneously in one pulse train. The shorter pulse duration with respect to the duration obtained
with the laser shown in \autoref{fig:ion-yield} is attributed to the 80~bar stagnation pressure
required to produce pyrrole-water clusters at high density.

The inset in \autoref{fig:tof} shows a zoomed-in TOF-MS spectrum. The complete, integrated time
trace was split into individual traces to obtain a TOF spectrum for each specific x-ray pulse.
Afterward, the TOF spectrum corresponding to a specific x-ray pulse was converted to the
mass-over-charge ratio in units of $u/e$. The main ion peaks of the TOF-MS spectrum are labeled and
attributed to the pyrrole-water parent ion (PW$^+$), pyrrole-water fragment ions, and remaining seed
gas contributions (He$^+$). A detailed analysis of the TOF-MS spectrum is beyond the scope of this
paper and will be discussed in a future publication. However, all ion signals induced by the x-rays
can be clearly resolved in the TOF-MS spectrum, demonstrating that the COMO setup is fully
compatible with the chosen burst mode operation at the EuXFEL. This will significantly increase the
data-acquisition duty cycle, and thus the utilization of x-ray pulses, compared to experiments using
standard short-pulse valves up to a factor of $\ordsim4$. It should be mentioned here that there are
no relevant restrictions on the number of pulses with which the valve is operated, up to several
hundred pulses per second. By modifying the electric driver longer molecular pulses will be possible
to cover the 400~\us duration of EuXFEL pulse trains. Furthermore, our results demonstrate that the
obtained molecular sample densities are clearly sufficient to conduct experiments with x-rays at
free-electron lasers using COMO. In addition, exploiting the electrostatic deflector, the x-ray
induced helium-ion signal, from the molecular beam seed gas, is highly suppressed. However, to
implement 100~\us or longer molecular beam pulses under conditions to produce significant densities
of molecular clusters, further optimization of the valve performance is required.

\section{Conclusion}
In summary, the permanently available molecular-beam injector setup COMO was developed, set up, and
commissioned using TOF-MS experiments both in-house at CFEL with optical-laser-ionization and at
EuXFEL with x-ray-ionization detection. The performance of the long-pulse-valve molecular-beam
source was characterized and proven to be compatible with the burst mode operation of the EuXFEL.
Our results show that the COMO setup can generate cold ($T\sim1$~K) and dense
($\rho\sim10^8$~cm$^{-3}$) molecular-beam pulses with pulse durations on the order of 100~\us. The
high molecular beam density allows experiments with data recorded in single-shot mode. In general,
the COMO molecular beam source can be combined with various electron- and ion-spectrometers, along
with large-area x-ray detectors. With the included electrostatic deflector, pure state-, size-, and
isomer-selected samples of polar molecules and clusters can be provided to the interaction region.

Hence, COMO proves to be an excellent source for various experiments at the EuXFEL, such as the
recording of molecular movies using ion, electron, and x-ray imaging. This enables a diverse range
of science from atomic, molecular, and cluster physics to materials and energy science, as well as
chemistry and biology.

\section{Acknowledgments}
We gratefully acknowledge Uzi Even, Nachum A.\ Lavie, and Ronny Barnea for designing and making
commercially available the long-pulse Even-Lavie valve and for sharing early test results from their
laboratory.

We acknowledge financial support by Deutsches Elektronen-Synchtrotron DESY, a member of the
Helmholtz Association (HGF), the Clusters of Excellence ``Center for Ultrafast Imaging'' (CUI,
EXC~1074, ID~194651731) and ``Advanced Imaging of Matter'' (AIM, EXC~2056, ID~390715994) of the
Deutsche Forschungsgemeinschaft (DFG), by the European Research Council under the European Union's
Seventh Framework Program (FP7/2007-2013) through the Consolidator Grant COMOTION (614507), the
European Union’s Horizon 2020 research and innovation programme under the Marie Sklodowska-Curie
Grant Agreement ``Molecular Electron Dynamics investigated by Intense Fields and Attosecond Pulses''
(MEDEA, 641789), the Helmholtz Foundation through funds from the Helmholtz-Lund International
Graduate School (HELIOS, Helmholtz project number HIRS-0018), and from the German Federal Ministry
of Education and Research (BMBF) and the Swedish Research Council (VR~2021-05992) through the
Röntgen-Ångström cluster (project ``Ultrafast dynamics in intermolecular energy transfer: elementary
processes in aerosols and liquid chemistry'' UDIET, 05K22GUA). We also acknowledge the scientific
exchange of the Center for Molecular Water Science (CMWS). For the experiments at EuXFEL we
acknowledge EuXFEL in Schenefeld, Germany, for the provision of XFEL beam time at the SQS instrument
and thank the staff for their assistance.

L.H.\ acknowledges the support by the National Natural Science Foundation of China (11704147 and
92261201) and a fellowships within the framework of the Helmholtz-OCPC postdoctoral exchange
program. F.T.\ acknowledges funding by the Deutsche Forschungsgemeinschaft (DFG, project 509471550,
Emmy-Noether programme) and by the MaxWater initiative of the Max-Planck-Gesellschaft. S.B.\ and
L.S.\ acknowledge funding by the Helmholtz Initiative and Networking Fund.

\section{Data availability}
The data recorded for the experiment at the EuXFEL are available at
\href{https://doi.org/10.22003/XFEL.EU-DATA-002388-00}{https://doi.org/10.22003/XFEL.EU-DATA-002388-00}.
The data recorded in the laser lab at DESY are available from the corresponding author upon
reasonable request.

\section{Code availability}
The code for the data analysis is available from the corresponding author upon reasonable request.

\bibliography{string,cmi}
\onecolumngrid
\clearpage
\listofnotes
\end{document}